\input{aipcheck}

\documentclass[
    ,final            
  ]
  {aipproc}

\layoutstyle{6x9}
\def\beq{\begin{equation}}
\def\eeq{\end{equation}}

\def\beqn{\begin{eqnarray}}
\def\eeqn{\end{eqnarray}}

\begin{document}

\title{Introduction to Electroweak Symmetry Breaking\footnote{Lectures given at the XIII Mexican School of Particles and Fields, 2-11 October, 2008, Sonora, Mexico.} }


\author{S. Dawson}{
  address={Physics Department, Brookhaven National Laboratory \\
Upton, New York~~11973}
}

\begin{abstract}

In these lectures, I review the status of the electroweak 
sector of the Standard Model, 
with an emphasis on the importance of radiative corrections 
and searches for the
Standard Model Higgs boson.  A discussion of the special role of the TeV energy
scale in electroweak physics is included. 

\end{abstract}

\maketitle

\section{Introduction}

The Standard Model (SM) is the backbone of elementary particle
physics--not only does it provide a consistent framework for studying the
interactions of quark and leptons, but it also gives predictions which have
been extensively tested experimentally.  
In these notes, I review the electroweak
sector of the Standard Model, discuss the calculation of electroweak 
radiative corrections to observables, and 
summarize the status of SM Higgs boson searches.

Despite the impressive experimental successes, however,
the electroweak theory is not
completely satisfactory and the mechanism of electroweak symmetry 
breaking is untested. I will discuss the logic behind the oft-repeated
statement: ``There must be new physics at the $TeV$ scale''. These
lectures reflect my strongly held belief that upcoming results from 
the LHC will fundamentally change our understanding of electroweak
symmetry breaking.

\section{The Standard Model}

The electroweak sector of the Standard Model has been reviewed extensively
in the 
literature\cite{Djouadi:2005gi,Quigg:2007fj,Gunion:1989we,Peskin:1995ev,Rainwater:2007cp,Quigg:1983gw,Dawson:1998yi,Reina:2005ae} and I provide only a brief summary here. 
The Weinberg- Salam model is an $SU(2)_L \times U(1)_Y$ gauge theory containing
three $SU(2)_L$ gauge bosons, $W_\mu^i$, $i=1,2,3$, and one $U(1)_Y$
gauge boson, $B_\mu$, with  kinetic energy terms,
\beq
{\cal L}_{\rm KE} =-{1\over 4}\Sigma_{i=1}^3 W_{\mu\nu}^i W^{\mu\nu i}
-{1\over 4} B_{\mu\nu} B^{\mu\nu}\,  ,
\eeq
where
\beqn
W_{\mu\nu}^i&=& \partial_\nu W_\mu^i-\partial _\mu W_\nu^i
-g \epsilon^{ijk}W_\mu^j W_\nu^k 
\nonumber \\
B_{\mu\nu}&=&\partial_\nu B_\mu-\partial_\mu B_\nu\, .
\eeqn
A mass term for the $W$ and $B$ gauge bosons would break the $SU(2)_L\times U(1)$
gauge symmetry. 

Coupled to the gauge fields is a complex scalar $SU(2)_L$
doublet, $\Phi$,
\beq
\Phi
= \left(\begin{array}{c}
 \phi^+   \\
 \phi^0   \end{array}\right) \, ,
\eeq
with a  scalar potential given by
\beq
 V(\Phi)=\mu^2 \mid \Phi^\dagger\Phi\mid +\lambda
\biggl(\mid \Phi^\dagger \Phi\mid\biggr)^2\,  ,
\label{wspot}
\eeq
($\lambda>0$).
This is the most general renormalizable and $SU(2)_L$ invariant
potential.

The state of minimum
energy for $\mu^2<0$ is not at $\phi^0=0$ and hence
the scalar field develops
a vacuum expectation value (VEV).
The direction of the minimum in $SU(2)_L$ space is not determined
since the potential depends only on 
the combination 
$\Phi^\dagger \Phi$
and we arbitrarily choose
\beq
\langle \Phi\rangle
= {1\over\sqrt{2}} \left(\begin{array}{c}
 0   \\
 v   \end{array}\right)\, .
\label{vevdef}
\eeq
With this choice the scalar doublet has $U(1)_Y$ charge
(hypercharge) $Y_\Phi=1$ and the
 electromagnetic charge is\footnote{The $\tau_i$ are
 the Pauli matrices with $Tr(\tau_i\tau_j)
=2\delta_{ij}$.}
\beq
Q_{em}={(\tau_3 +Y)\over 2} \, ,
\label{qdef}
\eeq
yielding an unbroken electromagnetic charge symmetry:
\beq 
Q_{em} \langle \Phi\rangle
= 0\, .
\eeq

\begin{table}
\begin{tabular}{lccr}
\hline
   \tablehead{1}{r}{b}{Field}
  & \tablehead{1}{r}{b}{$SU(3)$}
  & \tablehead{1}{r}{b}{$SU(2)_L$}
  & \tablehead{1}{r}{b}{$U(1)_Y$}   \\
\hline
$Q_L=\left(\begin{array}{c}
u_L\\ d_L \end{array}\right)$
   &    $3$          & $2$&  $~{1\over 3}$
\\
$u_R$ & $3$ & $1$& ${4\over 3}$
\\
$ d_R$ & $ 3$ & $1$&  $~-{2\over 3}$
\\
$L_L=\left(\begin{array}{c}
\nu_L\\ e_L \end{array}\right)
$  & $1$             & $2$& $~-1$
\\
$e_R$ & $1$             & $1$& $~-2$ 
\\
$\Phi= \left(\begin{array}{c}
\phi^+\\ \phi^0 \end{array}\right)
$  & $1$             & $2$& $1$ 
\\
\hline
\end{tabular}
\caption{Standard Model Particles}
\label{particles}
\end{table}

The
contribution of the scalar doublet to the Lagrangian is,
\beq
{\cal L}_s=(D^\mu \Phi)^\dagger (D_\mu \Phi)-V(\Phi)\quad ,
\label{scalepot}
\eeq
where
\beq
D_\mu=\partial_\mu -i {g\over 2}\tau\cdot W_\mu-i{g^\prime\over 2}
B_\mu Y.
\eeq
In unitary gauge, the scalar doublet can be written  in terms
of a physical scalar Higgs field, $h$, as
\beq
\Phi={1\over \sqrt{2}}\left(\begin{array}{c}  0 \\
 v+h\end{array}\right)                  \, ,
\eeq
which gives the contribution to the gauge boson masses
from the scalar kinetic energy term of Eq. \ref{scalepot},
\beq
M^2_{Gauge~Boson}\sim
{1\over 2} (0 ,v )
\biggl({1\over 2}g \tau\cdot W_\mu
+{1\over 2} g^\prime B_\mu
\biggr)^2 \left(\begin{array}{c}  0 \\  v \end{array}
\right).
\label{gaugem}
\eeq
The physical gauge fields are  two
charged fields, $W^\pm$, and two  neutral gauge
bosons, $Z$ and $A$,
\beqn
W^{\pm}_\mu&=&
{1\over \sqrt{2}}(W_\mu^1 \mp i W_\mu^2)\nonumber \\
Z_\mu&=& {-g^\prime B_\mu+ g W_\mu^3\over \sqrt{g^2+g^{\prime~2}}}
\nonumber \\
A_\mu&=& {g B_\mu+ g^{\prime} W_\mu^3\over \sqrt{g^2+g^{\prime~2}}}.
\label{masseig}
\eeqn
The gauge bosons   obtain  masses
via Eq. \ref{gaugem}:
\beqn
M_W^2 &=& {1\over 4} g^2 v^2\nonumber \\
M_Z^2 &=& {1\over 4} (g^2 + g^{\prime~2})v^2\nonumber \\
M_A& = & 0.
\eeqn
Three of the degrees of freedom of the complex scalar doublet have been absorbed by
the gauge bosons to generate longitudinal polarizations for the $W$ and $Z$
gauge bosons.  This is the Higgs mechanism. 

Since the massless photon must couple with electromagnetic
strength, $e$, the coupling constants 
define a  weak mixing angle $\theta_W$,
\beqn
e&=& g \sin\theta_W \nonumber \\
e&=& g^\prime \cos\theta_W
\,  .
\label{sdef}
\eeqn

\begin{figure}[t]
\includegraphics[bb=28 48 537 455,scale=0.40]{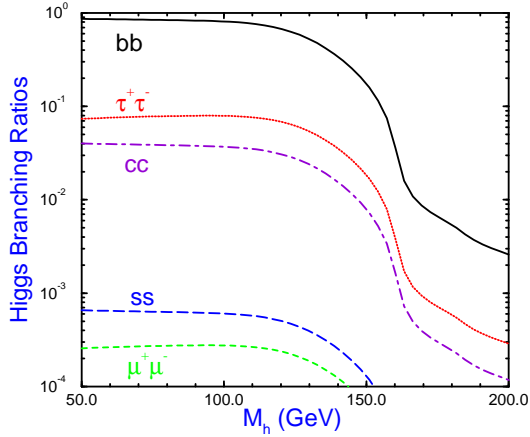} 
\caption{Standard Model Higgs boson  branching ratios into fermion/anti-fermion
 pairs.}
\label{fg:higgsbrs1}
\end{figure}

Fermions can easily be included in the theory and we consider
the electron and its neutrino as an example.  
It is convenient to write the fermions in terms of their left-
and right-handed projections,
\beq
\psi_{L,R}={1\over 2}(1\mp\gamma_5)\psi
\,\, .
\eeq
The lef-handed fermions
are assumed to transform as an
$SU(2)_L$ doublet, 
\beq
L_L=
\left(\begin{array}{c}
\nu_L\\ e_L \end{array}\right)\, .
\eeq
From Eq. \ref{qdef}, the hypercharge of the lepton doublet
must be $Y_L=-1$. 
Since the neutrino is (at least approximately) massless, it can have only
one helicity state which is taken to be $\nu_L$. 
From the four-Fermi theory of weak interactions, we know that the
$W$-boson couples only to left-handed fermions (see
for example, Ref. \cite{Quigg:1983gw}).
 Experimentally,
the
right-handed fields do not interact
with the $W$ boson, and so the right-handed electron, $e_R$, must be an
$SU(2)_L$ singlet and  so has $Y_{e_R}=-2$.  Using these hypercharge
assignments, the leptons can be coupled
in a gauge invariant manner
 to the $SU(2)_L\times
U(1)_Y$ gauge fields,
\beq
{\cal L}_{lepton}
=i {\overline e}_R \gamma^\mu
\biggl(\partial_\mu-i{g^\prime\over 2}Y_e B_\mu\biggr)e_R+
i{\overline L}_L\gamma^\mu
\biggl(\partial_\mu-i {g\over 2}\tau\cdot W_\mu
-i{g^\prime\over 2}Y_L
B_\mu\biggr)L_L\,\,.
\eeq

All of the known fermions can be accommodated in the Standard
Model in a similar manner.
The $SU(2)_L$ and $U(1)_Y$ charge assignments of
the first generation of fermions 
are given in Table 1.  The left-handed fermions are $SU(2)_L$ doublets,
while the right-handed fermions are $SU(2)_L$ singlets.  The $SU(3)$
color charge assignments are also listed for convenience.
\begin{figure}[t]
\includegraphics[bb=18 18 552 482,scale=0.40]{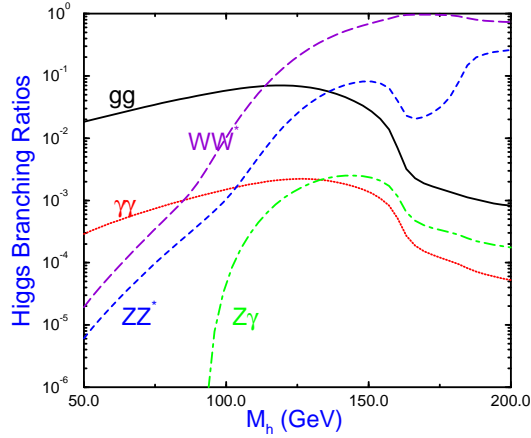}
\caption{Standard Model Higgs boson  branching ratios into gauge
boson pairs.}
\label{fg:higgsbrs2}
\end{figure}

The parameter $v$ can be found  from 
the charged current for $\mu$ decay,
$\mu\rightarrow e {\overline \nu}_e \nu_\mu$, by making
the identification,
\begin{equation}
{G_\mu\over \sqrt{2}}={g^2\over 8 M_W^2}
\, .
\end{equation} 
The interaction strength for muon decay is measured
very accurately to be $G_\mu=1.16637\times 10^{-5}~GeV^{-2}$ and
can be used to determine $v=(\sqrt{2} G_\mu)^{-1/2} = 246~GeV$.

A fermion mass term takes the form
\beq
{\cal L}_{mass}=-m{\overline{\psi}}\psi=-m
\biggl({\overline{\psi}}_L\psi_R+
{\overline{\psi}}_R\psi_L\biggr) 
\,\, .
\label{massterm}
\eeq
As is obvious from Table 1, the left-and right-handed
fermions transform differently under $SU(2)_L$ and $U(1)_Y$ and
electroweak 
gauge invariance therefore forbids a term of the form of
 Eq. \ref{massterm}.  
 The Higgs
boson can be used to give the fermions mass, however. The
gauge invariant  Yukawa coupling of the
Higgs boson to the up and down quarks  is
\beq
{\cal L}_d=-\lambda_d {\overline Q}_L \Phi d_R + h.c.\,  .
\eeq
This gives the effective coupling,
\beq
-\lambda_d {1\over\sqrt{2}}
({\overline u}_L,~ {\overline d}_L)\left(
\begin{array}{c}  0 \\
v+ h \end{array} \right) d_R + h.c.\, ,
\eeq
which can be seen to yield a mass term for the down quark if
we make the identification
\beq 
\lambda_d = {m_d \sqrt{2}\over v}.
\eeq
In order to generate a mass term for the up quark note that
$\Phi^c \equiv - i \tau_2 \Phi^*$ is an $SU(2)_L$
invariant, allowing the coupling,
\beq
{\cal L}_u=-\lambda_u {\overline Q}_L \Phi^c u_R + h.c. \, ,
\eeq
which generates a mass term for the up quark.  Similar couplings
can be used to generate mass terms for the charged leptons.
Since the neutrino has no right handed partner, it remains 
massless.  Hence a single scalar Higgs doublet not only generates masses for 
the gauge bosons, but also for fermions.  Unfortunately, the size
of the fermion masses remains unexplained. 

The couplings of the Higgs boson to the fermions and gauge bosons
are directly proportional to their masses (by construction), 
which has the implication that the Higgs boson decays primarily
to the heaviest particles kinematically allowed.  At tree level, the
Higgs couplings to photons and gluons vanish since the photon
and gluon are massless. These couplings first
arise at $1$-loop and hence are sensitive to new non-SM  particles which
may propagate in the loops. The Higgs boson  branching ratios are 
shown in Figs. \ref{fg:higgsbrs1} and \ref{fg:higgsbrs2}
 and they can easily  be calculated including
higher order corrections using the programs 
HDECAY\cite{hdecay} or FEYNHIGGS\cite{Frank:2006yh}.

One of the most important points about the Higgs mechanism is
that all of the couplings of the Higgs boson to fermions and
gauge bosons are completely determined in terms of gauge coupling
constants and fermion masses. The potential of Eq. \ref{wspot}
has two free parameters, $\mu$ and $\lambda$.  We can trade
these for
\beqn
v^2&=&-{\mu^2\over 2 \lambda}= (246~GeV)^2\nonumber \\
M_h^2&=& 2 v^2 \lambda.
\label{mhdef}
\eeqn
There are no remaining adjustable
parameters and so Higgs production and decay processes
can be computed unambiguously in terms of the Higgs mass alone,
making the Higgs sector of the theory completely determined
in the SM.

When the scalar potential is expressed in terms of $v$ and $M_h$, it
becomes,
\begin{equation}
V={M_h^2\over 2}h^2 +{M_h^2\over 2 v}h^3+{M_h^2\over 8 v^2}h^4
\, ,
\end{equation}
and it is apparent that for heavy Higgs masses ($M_h\sim 1~TeV$), the Higgs
self-interactions become strong. 

\section{Experimental Searches for the Higgs Boson}
\subsection{LEP}
\begin{figure}[t]
\includegraphics[scale=0.40,angle=-90]{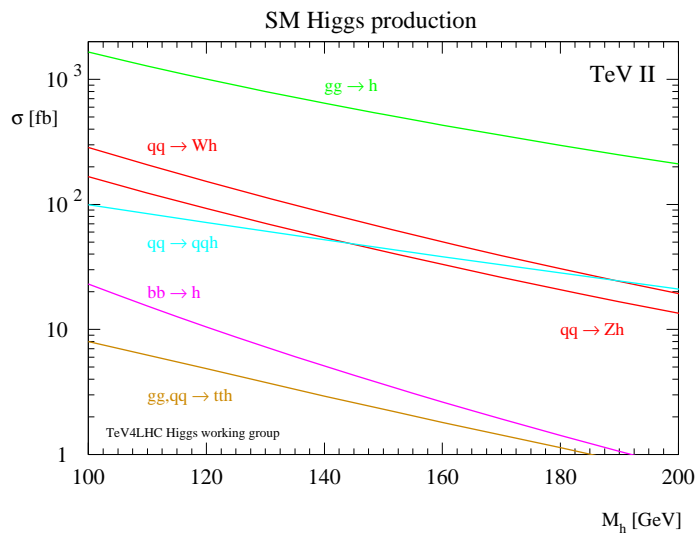}
\caption{Cross sections for SM Higgs boson production
processes at the Tevatron ($\sqrt{s}\!=\!2$~TeV), including higher
order QCD corrections. From
Ref.~\cite{Aglietti:2006ne}.\label{fig:sm_higgs_tev}}
\end{figure}

The Higgs boson was directly searched for at the LEP collider through the
process $e^+e^-\rightarrow Zh$ at energies up to $\sqrt{s}=209~GeV$.
  The Higgs boson decays to the heaviest
particles kinematically accessible ($b{\overline b}$ and $\tau^+\tau^-$
for the LEP searches) and the $Z$ decays roughly $70\%$ of the
time to jets, $20\%$ to neutrinos, and $10\%$ to charged leptons.
 The LEP
experiments searched in all of these channels
and obtained the limit on a SM Higgs boson\cite{Barate:2003sz},
\begin{equation}
M_h>114.4~GeV
\, .
\label{leplim}
\end{equation}
This limit can potentially
be
evaded by
constructing models where the Higgs boson decays to non-SM invisible
particles with large  branching ratios or the Higgs has highly suppressed
non-SM couplings
to the $Z$\cite{Chang:2008cw}. A Higgs boson with couplings an order
of magnitude smaller than the SM $hZZ$ coupling has been ruled out by the
LEP experiments for $M_h < 80~GeV$.

\subsection{Tevatron}

\begin{figure}[t]
  \includegraphics[height=.1\textheight]{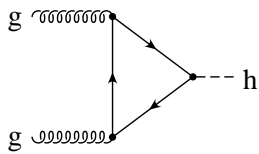}
  \caption{Higgs production from gluon fusion. The dominant contribution
is from a top quark loop.}
\label{fg:ggh}
\end{figure}

\begin{figure}[t]
  \includegraphics[height=.4\textheight]{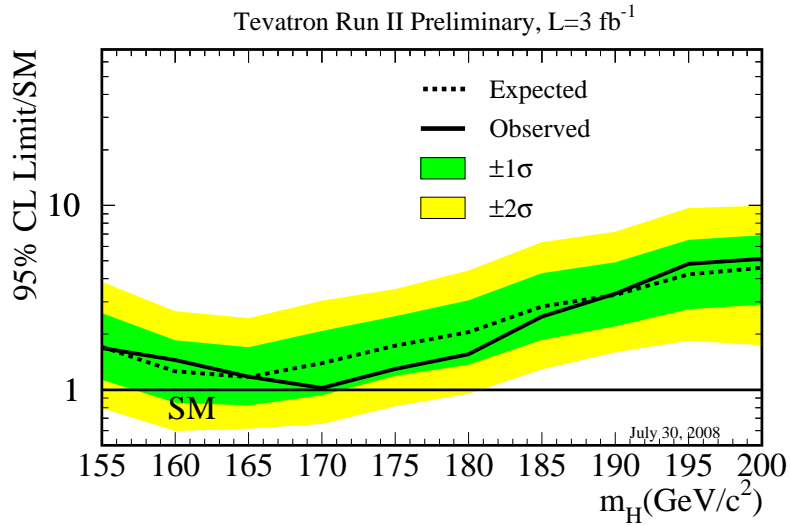}
  \caption{Tevatron exclusion of a Higgs boson with
$M_h=170~GeV$. From Ref. \cite{Herndon:2008uv}.}
\label{fg:tevexc}
\end{figure}

The production rates for the SM Higgs boson at the Tevatron
are shown in Fig. \ref{fig:sm_higgs_tev}.  The largest rate is that
for the partonic subprocess $gg\rightarrow h$ (Fig. \ref{fg:ggh}). For
$M_h< 140~GeV$, the Higgs decays almost entirely to $b {\overline b}$
pairs, as seen from Fig. \ref{fg:higgsbrs1}.
  Unfortunately, the $b {\overline b}$ background
is many orders of magnitude larger than the signal and it does not appear
possible to extract a Higgs signal from the $gg\rightarrow
h\rightarrow b {\overline b}$ channel\cite{Rainwater:2007cp}. 
For $M_h> 140~GeV$, however,
the rate to $WW^*$\footnote{$W^*$ denotes a virtual $W$ and the branching
ratio of Fig. \ref{fg:higgsbrs1} implies a factor of the branching
ratio of $W^*\rightarrow {\overline f} f^\prime$.} grows with increasing Higgs boson mass
(see Fig. \ref{fg:higgsbrs2})
 and becomes
large near the $W^+W^-$ threshold.  Using this channel, the Tevatron experiments
have excluded a single point, $M_h=170~GeV$\cite{Herndon:2008uv}, as shown in Fig.
\ref{fg:tevexc}.

\begin{figure}[t]
  \includegraphics[height=.12\textheight]{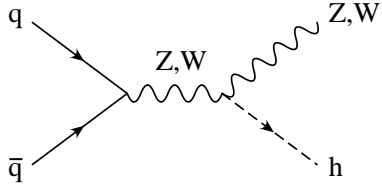}
  \caption{Higgs production in association with a vector boson.}
\label{fg:assp}
\end{figure}

\begin{figure}
\includegraphics[scale=0.55]{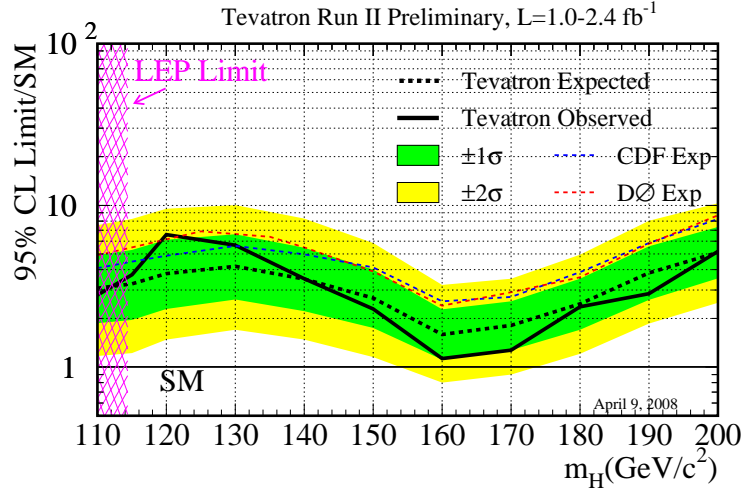}
\caption{$95\%$ exclusion limit for  a SM Higgs boson from the
Tevatron experiments. From
Ref.~\cite{Group:2008ds}.}
\label{fg:tevnew}
\end{figure}

The search for a relatively light Higgs boson, $M_h\sim 114~GeV$, 
proceeds at the Tevatron primarily through the associated production channel shown
in Fig. \ref{fg:assp}.  Although the cross section is smaller
than in the gluon fusion channel, the decay products of the $Z$ or
$W$ can be tagged and used to reduce the background.  The Tevatron
limits are normalized to the SM expectations and for $114~ GeV < M_h
< 170~GeV$ are between a factor of 3-7 above the SM predictions
as can be 
seen in Fig. \ref{fg:tevnew}\cite{Group:2008ds}.  
It is interesting to note that more than 70
different channels are used to obtain Fig. \ref{fg:tevnew}, making the statistical
combination of individual limits quite complicated.    The Tevatron exclusion 
results are rate limited and are expected to improve with
increasing luminosity.

\begin{figure}[t]
  \includegraphics[height=.4\textheight]{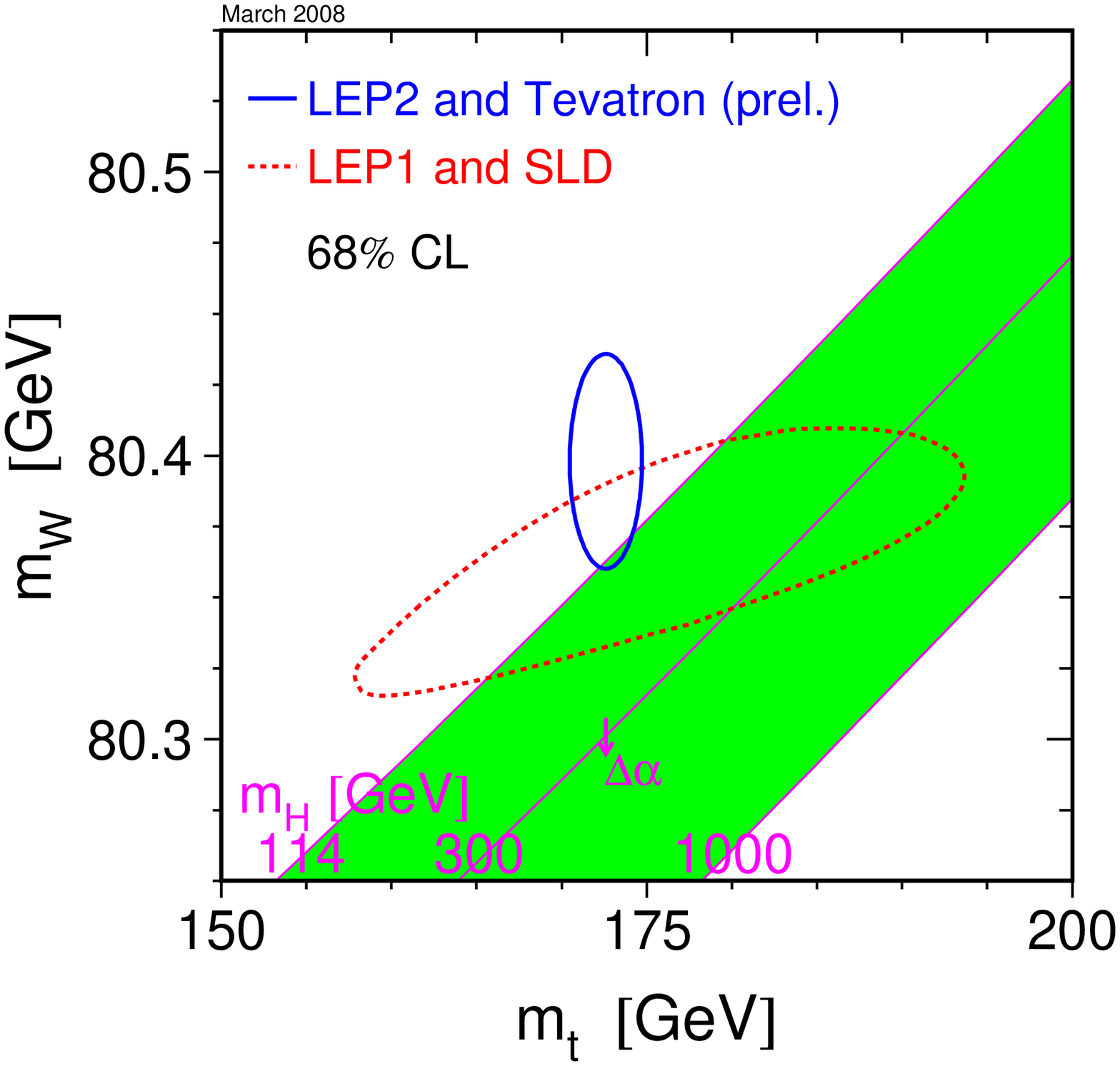}
  \caption{Comparison of the direct search results from LEP2 and the Tevatron for
$M_w$ and $m_t$ with those inferred from the consistency of the SM. From Ref. \cite{lepewwg}.
}
\label{mhlim}
\end{figure}

\begin{figure}[t]
  \includegraphics[height=.4\textheight]{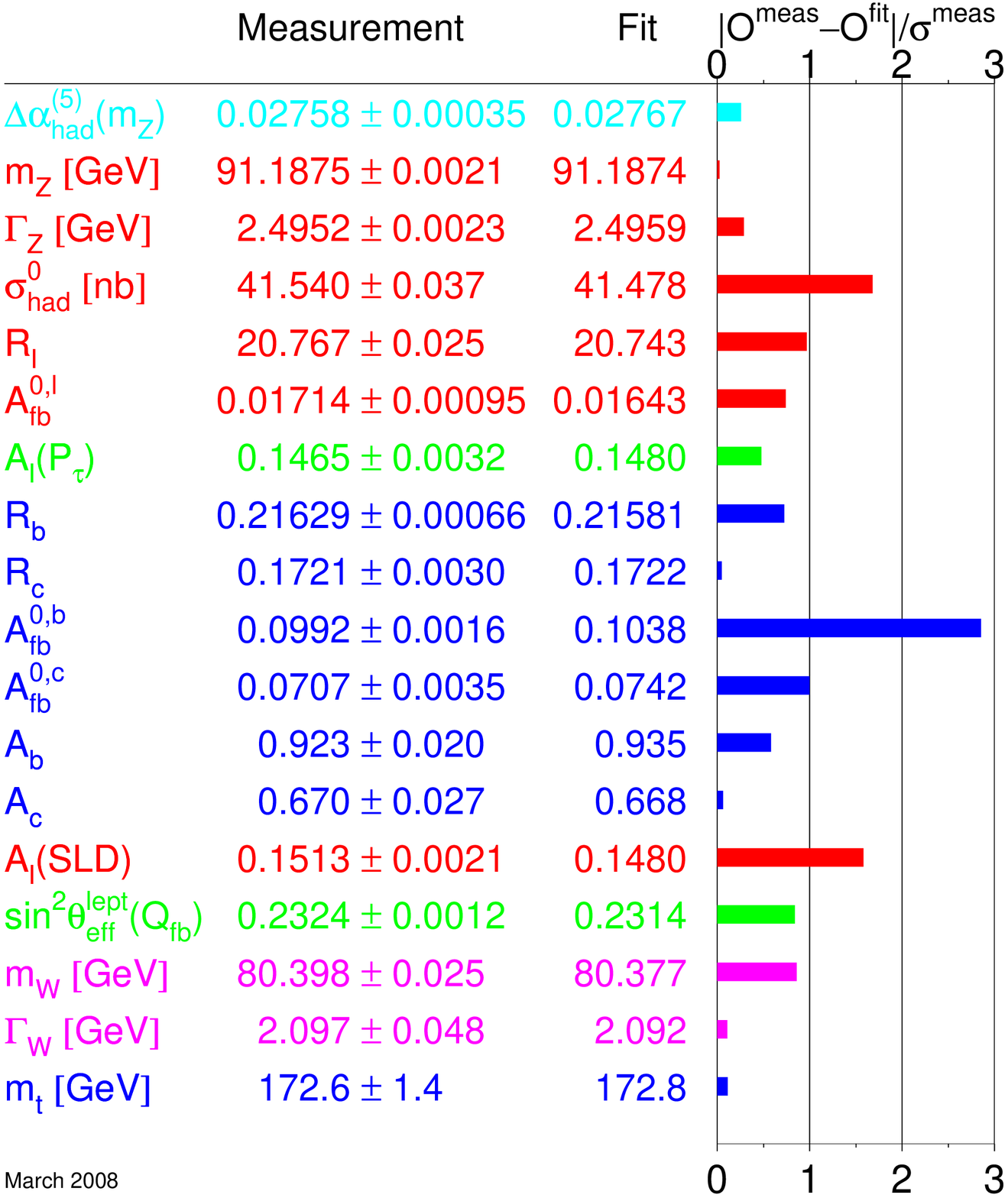}
  \caption{Comparison of electroweak measurements with a best fit
to the SM theory. From Ref. \cite{lepewwg}.}
\label{pullfig}
\end{figure}

\section{Limits from Precision Measurements}

The electroweak sector of the SM can be tested at the multi-loop
level due to its predictive nature.
In the electroweak sector of the SM, the gauge sector has four fundamental 
parameters, the $SU(2)_{L} \times U(1)_{Y}$ gauge coupling constants, 
$g$ and $g^\prime$, as well as the two parameters of the Higgs potential,
which are usually taken to be the  vacuum expectation value of the 
Higgs boson,  $v$, and the Higgs mass, $M_h$. Once these  parameters are fixed, 
all other physical quantities in the electroweak sector can be derived in terms 
of them (and of course the fermion masses and
CKM mixing parameters, along with the strong coupling constant 
$\alpha_s$).
Equivalently, the muon decay constant, $G_{\mu}$, 
the Z-boson mass, $M_{Z}$, and the fine structure constant,
 $\alpha$, can be used as input parameters. 
Experimentally, the measured values for these input parameters 
are\cite{Amsler:2008zz},
\begin{eqnarray}
G_{\mu} & = & 1.16637(1) \times 10^{-5} \; GeV^{-2}\\
M_{Z} & = & 91.1876(21) \; GeV\\
\alpha^{-1} & = & 137.035 999 679 (94)\, .
\end{eqnarray}

The $W$ boson mass
is defined through muon decay,  
\begin{equation}
M_W^2 = \frac{\pi \alpha}{\sqrt{2} G_\mu s_{\theta}^{2}}
\, .
\label{wdef}
\end{equation}
The SM satisfies $\rho =1$ at tree level and predicts
the weak mixing angle in terms of the gauge boson masses,
\begin{equation}
\rho = 1 =\frac{M_{W}^{2}}{M_{Z}^{2}c_{\theta}^{2}}\quad .
\label{rhodef}
\end{equation}
At tree level, all definitions of the weak mixing angle (Eq. \ref{sdef}) are equivalent, but the
definitions differ at higher order. 
In Eq. \ref{rhodef}, $M_W$ and $M_Z$ are the physical gauge boson 
masses, and this definition of the weak mixing angle, $s_\theta$,
corresponds to the on-shell scheme. 
Eqs. \ref{wdef} and \ref{rhodef} imply,
\begin{equation}
M_W^2={M_Z^2\over 2}\biggl\{ 1+\sqrt{1-{4\pi \alpha\over \sqrt{2}G_\mu M_Z^2}}
\biggr\}
\, .
\label{mwtree}
\end{equation}
At tree level, the SM therefore predicts from Eq. \ref{mwtree},
\begin{equation}
M_W(tree)=79.829~GeV\, ,
\end{equation}
in disagreement with the measured value\cite{lepewwg},
\begin{equation}
M_W(experiment)=80.399\pm0.025~GeV
\, .
\end{equation}

In order to obtain good agreement between theory and the experimental data, it
is crucial to include radiative corrections\cite{Hollik:1988ii,Jegerlehner:1991dq}.  
For example, the prediction 
for $M_W$ can be expressed as\cite{Sirlin:1980nh},
\begin{equation}
 M_{W}^{2} = \frac{\pi \alpha}{\sqrt{2}G_{\mu} s_{\theta}^{2}}
\biggl[1 + \Delta r_{SM} \biggr]\, ,
\label{rdef}
\end{equation}
where $\Delta r_{SM}$ summarizes the radiative corrections.
The dependence on the top quark mass, $m_t$, is particularly significant as
$\Delta r_{SM}$ depends on $m_{t}$ quadratically, 
\begin{eqnarray}
\Delta r_{SM}^t& = &
 - {G_\mu\over \sqrt{2}}{N_c\over 8 \pi^2}
\biggl( {c_\theta^2\over s_\theta^2}
\biggr) m_t^2+\log(m_t)~{\rm{terms}}\quad ,
\end{eqnarray}
where $N_c=3$ is the number of colors.

The top quark does not decouple from the theory even at energies far above
the top quark mass.  The decoupling theorem\cite{Appelquist:1974tg}
(which says that heavy particles do not affect low scale physics)
 is violated by the top
quark because the top quark couplings to both the Higgs boson and the longitudinal
components of the gauge bosons  are proportional to $m_t$ and  also because the
SM is not renormalizable without the top quark.  

The dependence of $M_W$ and other electroweak observables
on the Higgs boson mass is logarithmic and so predictions are much
less sensitive to $M_h$ than to $m_t$.
The complete contribution to $\Delta r_{SM}$ can be
approximated for a heavy Higgs\cite{Sirlin:1980nh},
\begin{equation}
\Delta r_{SM}=.070+\Delta r_{SM}^t+{\alpha\over \pi
s_\theta^2}{11\over 48}\biggl[ \log\biggl({M_h^2\over M_Z^2}\biggr)-{5\over 6}
\biggr]
+{\hbox{2-loop}}\, .
\label{drsm}
\end{equation}
The first term in Eq. \ref{drsm} results primarily from the scaling of $\delta \alpha$
from $q^2=0$ to $M_Z$.

The agreement between the prediction for the $W$ mass given by
Eqs. \ref{rdef} and \ref{drsm} with the measured value is a strong test
of the theory. The measurements of $M_W$ and $m_t$ can be used to infer
limits on the Higgs boson mass, as can be seen in Fig. \ref{mhlim}.  A relatively
light value of $M_h$ is clearly prefered. 
The measured values for some high energy observables from LEP, SLC, and the
Tevatron are listed in the left hand column of Fig. \ref{pullfig}.  The best
fit to the predictions of the SM (including radiative corrections) is given in
the right hand column.  The agreement between the data and the predictions is
compelling evidence for the validity of the SM at current energy scales.
From Fig. \ref{mhlim}, we see that the measured $W$ mass is slightly high
compared with the value extracted from the precision electroweak data 
of  Fig. \ref{pullfig}.

\begin{figure}[t]
  \includegraphics[height=.4\textheight]{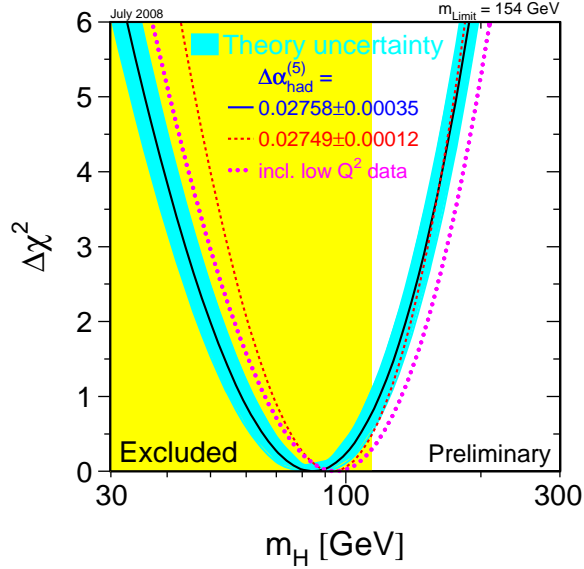}
  \caption{Best fit to the Higgs boson mass from data at LEP, SLC, and 
the Tevatron. From Ref. \cite{lepewwg}.}
\label{blue}
\end{figure}

The data can be used to obtain a prediction for the Higgs boson mass as seen
in the ``blue-band'' plot of Fig. \ref{blue}.  When the direct Higgs search results
from LEP and the Tevatron
and the results from low energy experiments such as atomic parity violation
are omitted, the best fit is\cite{lepewwg},
\begin{equation}
M_h=84^{+34}_{-26}~GeV\, .
\label{hfit}
\end{equation}
It is somewhat distressing that the best fit from the observables of Fig. \ref{pullfig}
is in the region excluded by the LEP direct search, Eq. \ref{leplim}.  
When the direct search results from LEP
are included, a $95\% $ confidence level upper bound is found,
\begin{equation}
M_h < 185~GeV\, .
\label{smlim}
\end{equation}
This limit assumes that there are no particles which contribute to the radiative
corrections to the SM observables of Fig. \ref{pullfig} beyond the SM particles
(and also that all couplings have their SM values). 
It is quite easy to evade the limit of Eq. \ref{smlim}
 in extentions of the SM. 

\begin{figure}[t]
  \includegraphics[height=.3\textheight]{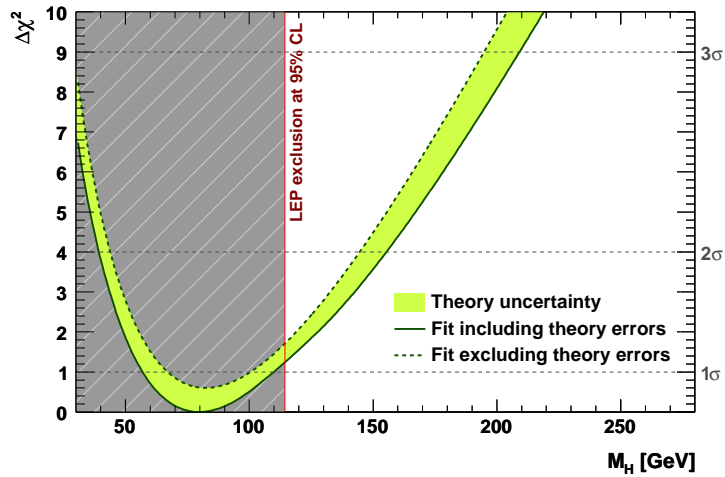}
  \caption{Blue-band plot from the GFITTER collaboration including an estimate of
theory errors. From Ref. \cite{Flaecher:2008zq}.}
\label{gblue}
\end{figure}

\begin{figure}[t]
  \includegraphics[height=.3\textheight]{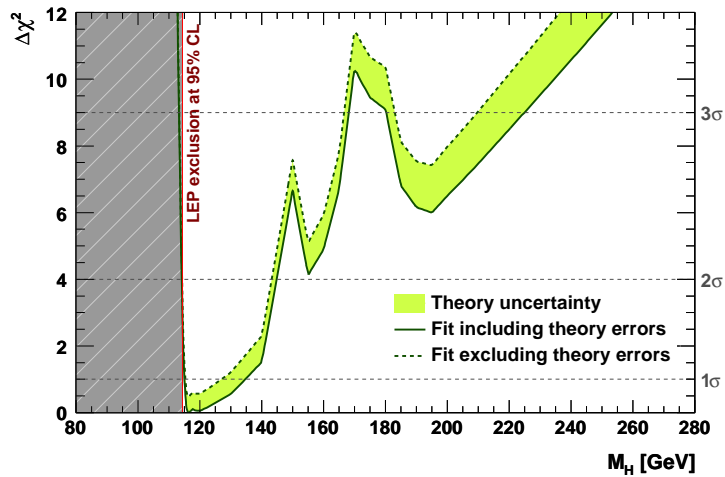}
  \caption{Blue-band plot from the GFITTER collaboration including the direct
search limits from LEP and from the Tevatron experiments with $3~fb^{-1}$
of data\cite{Flaecher:2008zq}.}
\label{bluetev}
\end{figure}

The fits shown in Figs. \ref{mhlim},~\ref{pullfig}
and \ref{blue} were performed by the LEP Electroweak Working Group.  Other
groups have also done global fits with slightly different methodologies and 
assumptions\cite{Flaecher:2008zq,Erler:2008ek}.  The GFITTER collaboration has
a version of the blue-band plot shown in Fig. \ref{gblue} which includes theory
uncertainties, which can be seen to be significant.  
The best fit to the Higgs mass from the GFITTER collaboration is,
\begin{equation}
M_h=80^{+30}_{-23}  GeV
\, ,
\end{equation}
in good agreement with the LEP Electroweak Working Group fit of Eq. \ref{hfit}.
The GFITTER result including the direct
search results  from both LEP and from $3~fb^{-1}$ of data at the Tevatron (Figs. \ref{fg:tevexc}
and \ref{fg:tevnew}) is shown in Fig.
\ref{bluetev}. It is interesting that the Tevatron search results are beginning
to influence the global fit to the Higgs mass.

\section{Why is the Standard Model Unsatisfactory?}
Although the SM is for the most part
 consistent with experimental data, most theorists
believe that it is incomplete.  In this section, I summarize the
arguments for the existence of physics beyond the SM.  

\subsection{Perturbativity and Triviality}
Theoretical bounds on the Higgs boson  mass have been  deduced on the grounds
of {\it triviality},
\cite{Lindner:1985uk,Chivukula:1999az,Hambye:1996wb} which can be summarized
as the requirement that the Higgs quartic coupling remain finite
at high energy scales. 
Consider the scalar sector of the SM,\footnote{$\mu^2<0$, $\lambda>0$.}
\beq
V(\Phi)=\mu^2\mid \Phi^\dagger \Phi\mid
 +\lambda (\mid \Phi^\dagger \Phi\mid )^2
\label{potsm}
\eeq
where the quartic coupling is
\beq
\lambda={M_h^2\over 2 v^2} \, .
\label{lamdef}
\eeq
The quartic coupling, $\lambda$,
changes with the effective energy scale $Q$ due to
the self interactions of the scalar field:
\beq
{d \lambda \over dt}={3 \lambda^2\over 4 \pi^2},
\label{lams}
\eeq
where $t\equiv \log(Q^2/Q_0^2)$ and $Q_0$ is some reference
scale.
Solving  Eq. \ref{lams},
\beqn
{1\over \lambda(Q)}&=&{1\over \lambda(Q_0)}-{3\over 4 \pi^2}
\log\biggl({Q^2\over Q^2_0}\biggr).
\eeqn
Summing the geometric series,
\beqn
\lambda(Q)&=& {\lambda(Q_0)\over 
\biggl[1-{3\lambda(Q_0)\over 4 \pi^2}\log({Q^2\over Q^2_0})
\biggr]}.
\label{lampol}
\eeqn
From Eq. ~\ref{lampol} we see that $\lambda(Q)$ blows up 
as $Q\rightarrow \infty$
 (called the Landau pole).  Regardless of how small $\lambda(Q_0)$
is, $\lambda(Q)$ will 
become infinite at some large value of $Q$.
Alternatively, $\lambda(Q_0)\rightarrow 0$ as
$Q\rightarrow 0$ with $\lambda(Q)>0$.  

The requirement that the quartic
coupling be finite at a high scale $\Lambda$,
\beq
{1\over \lambda(\Lambda)}>0, 
\eeq
can be interpreteted as a bound on the Higgs boson mass,
\beq
M_h^2 < {8 \pi^2 v^2\over 3 \log (\Lambda^2/v^2)}
\quad ,
\label{hlimL}
\eeq
(where we set $Q_0=v$).
Requiring that the SM be valid
up to the scale associated with grand unified models,  $\Lambda \sim 10^{16}~GeV$,
 yields
the approximate upper bound,
\beq
M_h < 160~GeV\, .
\eeq
As the scale $\Lambda$ becomes smaller, the limit on the Higgs
mass becomes progressively weaker. For large $\Lambda$, of course,
higher order and non-perturbative corrections must be 
included\cite{Sher:1988mj}. $\Lambda$ is often interpreted as the
scale of new physics, since above the scale $\Lambda$ the SM is
no longer a sensible theory. 

Another bound on the Higgs mass can be derived by the requirement that 
spontaneous symmetry breaking  occur,
\beq
V(v)< V(0).
\label{vacstab}
\eeq
This bound is essentially equivalent to the requirement
that $\lambda$ remain positive at all scales $\Lambda$.
(If $\lambda$ becomes negative, the potential is unbounded from
below and has no state of minimum energy.)
For small $\lambda$, the scaling is\cite{Gunion:1989we},
\beq
{d \lambda \over d t}={1\over 16 \pi^2}\biggl[
- 12 g_t^4+{3\over 16} (2 g^4 +(g^2+g^{\prime~2})^2)\biggr]
\,  ,
\label{renormsmall}
\eeq
where $g_t=m_t/v$ is the top quark Yukawa coupling. 
Eq. \ref{renormsmall} is easily solved to find,
\beq
\lambda(\Lambda)=\lambda(v)+{1\over 16 \pi^2}\biggl[
- 12 g_t^4+{3\over 16} (2 g^4 +(g^2+g^{\prime~2})^2)\biggr]
\log\biggl({\Lambda^2\over v^2}\biggr)
\, .
\eeq
Requiring
$\lambda(\Lambda)>0$ gives the bound on the Higgs boson mass,
\beq
M_h^2>{v^2\over 8 \pi^2}\biggl[
- 12 g_t^4+{3\over 16} (2 g^4 +(g^2+g^{\prime~2})^2)\biggr]
\log\biggl({\Lambda^2\over v^2}\biggr)
\, .
\eeq

\begin{figure}[t]
\includegraphics[scale=0.45,angle=90]{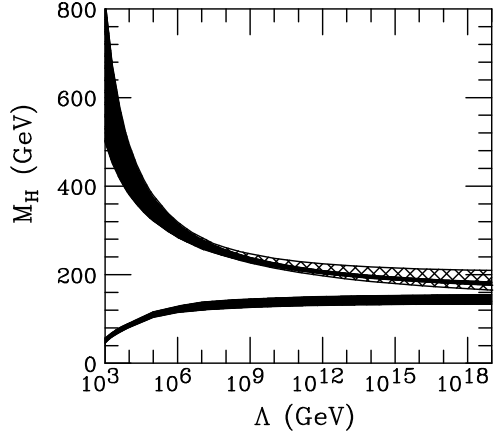}
\caption{Theoretical limits on the Higgs boson mass as a function of the
scale of new physics, $\Lambda$.  The
allowed region is between the curves. From
Ref.~\cite{Hambye:1996wb}.\label{fg:trivlim}}
\end{figure}

 A more careful 
analysis along the
same lines as above \cite{Sher:1988mj} using the $2$ loop renormalization group
improved effective potential\footnote{The renormalization
group improved effective potential sums all potentially large
logarithms, $\log(Q^2/v^2)$.} and the
running of all couplings gives the requirement 
that  if the Standard Model is valid
up to scales of order 
$10^{16}~ GeV$, then \cite{Sher:1988mj}
\beq
M_h(GeV)> 130 + 2(m_t-170)
\, .
\label{sherlim}
\eeq

\begin{figure}[t]
  \includegraphics[height=.12\textheight]{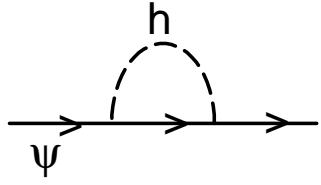}
  \caption{Fermion mass renormalization from an
internal Higgs boson.}
\label{fg:fey1}
\end{figure}
Eqs. \ref{hlimL} and \ref{sherlim} imply that if the SM is valid up to
around $10^{16}~GeV$, then the Higgs mass is restricted to be between
approximately $126~GeV$ and 
$160~GeV$\cite{Hambye:1996wb,Sher:1988mj,Isidori:2001bm,Kolda:2000wi}.  
It is interesting that this is precisely the region preferred by the
electroweak precision obervables of the previous section.
As the scale $\Lambda$ is reduced, the allowed 
range  for the Higgs mass
is enlarged. The theoretically allowed region for the Higgs mass
as a function of the scale $\Lambda$ is shown in 
Fig. \ref{fg:trivlim}.  It is important to remember that this bound 
assumes the SM with a single Higgs doublet.  In extentions of the SM
with extra Higgs doublets (for example, in 
supersymmetric models), it
is possible to evade the bound of Eq. \ref{sherlim}.

\subsection{Naturalness}

One of the most glaring theoretical inadequacies of the SM arises when
we compute quantum corrections to the Higgs boson mass.  One-loop corrections 
to the Higgs mass have the undesirable feature that they depend quadratically
on high scale physics\cite{Drees:1996ca}.
The basic point can be illustrated with a  simplified version of the SM 
containing 
 a single  fermion, $\psi$,
coupled to a massive  Higgs scalar, $\phi$,
\beq
{\cal L}_\phi=
{\overline \psi}(i \partial)\psi
+\mid \partial_\mu \phi\mid^2-m_S^2 \mid \phi\mid^2
-\biggl( {\lambda_F\over 2}
{\overline \psi}\psi \phi +{\rm h.c.} \biggr)  
\, .
\label{ephilag}
\eeq
Assume that this Lagrangian leads to spontaneous symmetry 
breaking and  $\phi=(h+v)/\sqrt{2}$, with $h$ a physical Higgs
boson.  
After spontaneous symmetry breaking,
the fermion acquires a mass, $m_F=\lambda_F v/\sqrt{2}$. 
Consider the fermion self-energy  arising from
the  scalar loop corresponding to Fig. \ref{fg:fey1}.
\beq
-i\Sigma_F(p)=\biggl({-i\lambda_F\over \sqrt{2}}\biggr)^2
(i)^2
\int{d^4k\over (2\pi)^4}{(k+m_F)\over [k^2-m_F^2][(k-p)^2-m_S^2]}
\quad .
\eeq
The renormalized fermion mass is $m_F^r=m_F+\delta m_F$, with
\beqn
\delta m_F &=& \Sigma_F(p)\mid_{ p=m_F}
\nonumber \\
&=& i {\lambda_F^2\over 32\pi^4}\int_0^1 dx \int  d^4 k^\prime
{m_F (1+x)\over [k^{\prime 2} -m_F^2 x^2-m_S^2(1-x)]^2}
\quad .
\label{meren}  
\eeqn
The integral can be performed in Euclidean space with
a momentum space cut-off using the fact that for a
symmetric integral\cite{Dawson:1997tz},
\beq \int d^4 k_E f(k_E^2)=\pi^2
\int^{\Lambda^2}_0 y dy f(y)
\quad .
\label{intfacs}  
\eeq 
In Eq. \ref{intfacs}, $\Lambda$ is a high energy cut-off, presumably of the
order of the Planck scale or a grand unified scale.
The renormalization of the fermion   mass is, 
\beqn
\delta m_F&=& -{\lambda_F^2 m_F\over 32\pi^2}\int_0^1~dx (1+x)
\int^{\Lambda^2} _0 {y dy\over [y+m_F^2x^2+m_S^2(1-x)]^2}
\nonumber \\
&=& -{3 \lambda_F^2 m_F\over 64\pi^2} \log\biggl({\Lambda^2
\over m_F^2}\biggr) + ....
\eeqn
where the $....$  indicates terms independent of the cutoff
or which vanish when $\Lambda\rightarrow\infty$.
This correction clearly corresponds to a well-defined expansion for $m_F$.

\begin{figure}[t]
  \includegraphics[height=.1\textheight]{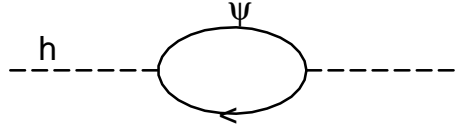}
  \caption{Fermion mass renormalization from a fermion loop, $\phi$.}
\label{fg:fey2}
\end{figure}
In the limit
in which the fermion mass vanishes,
Eq. \ref{ephilag} is invariant under the chiral transformations,
\beqn
\psi_L & \rightarrow  & e^{i\theta_L}\psi_L
\nonumber \\
\psi_R &\rightarrow & e^{i\theta_R}\psi_R,
\eeqn
and setting the fermion mass to zero increases the symmetry
of the theory.  
Since the Yukawa coupling (proportional to the fermion
 mass term) breaks
this symmetry, the corrections to the mass must be proportional to $m_F$. 

The situation is quite different, however, for the
renormalization of the scalar  mass  from
a fermion loop (Fig. \ref{fg:fey2})
 using the same Lagrangian (Eq. \ref{ephilag}),
\beq
-i\Sigma_S(p^2)=\biggl({-i\lambda_F\over\sqrt{2}}\biggr)^2
(i)^2 (-1)\int {d^4 k \over (2 \pi)^4}
{Tr[(k+m_F)((k-p)+m_F)]\over
(k^2-m_F^2)[(k-p)^2-m_F^2]}
\quad .
\eeq 
Integrating  as before with a momentum
space cutoff,
\beq
\delta M_h^2
=-{\lambda_F^2\over 8\pi^2}
\Lambda^2+
...
\label{quads}
\eeq
The Higgs boson mass depends {\bf {\it quadratically}} on the
high scale cut-off $\Lambda$.
Note that the correction is {\it not} proportional to $M_h$. 
Setting the Higgs mass equal to zero does not 
increase the symmetry of the Lagrangian and  there is nothing
that protects the Higgs mass from large corrections.

In the Standard Model, we expect that
the physical Higgs boson
 mass, $M_h$, is of the order of a few hundred $GeV$ from
the precision results discussed in the previous section.
The quadratic contributions to the Higgs boson mass renormalization in the SM
are\cite{Kolda:2000wi},
\begin{eqnarray}
\delta M_h^2&=& {3\over 8 \pi v^2}\Lambda^2
\biggl( 6 M_W^2+3 M_Z^2 +3 M_h^2-12 M_t^2\biggr)
\nonumber \\
&\sim & -\biggl({\Lambda\over .7~TeV}~200~GeV\biggr)^2
\, .
\label{quadh}
\end{eqnarray}
Eq. \ref{quadh} suggests that in order not to have large
cancellations, $\Lambda$ should be ${\cal O}(TeV)$.
This is known as the {\it{hierarchy~ problem}}:  Why should $\Lambda$ 
be ${\cal O}(TeV)$ and not the Planck scale?
Understanding the hierarchy problem as expressed by
Eq. \ref{quadh} has stimulated much model building.  The basic
approach is to postulate new particles which contribute to the
Higgs mass renormalization
 at one loop and cancel the SM contributions.  Supersymmetric
models do this by postulating scalar particles associated with the known 
fermions with just the right couplings to cancel the SM
contributions to Eq. \ref{quadh}, while Little Higgs type models
cancel the SM quadratic contributions using particles with the same
spin as the SM particles. In both cases, the models contain $TeV$
scale particles which can potentially be observed at the LHC.

\subsection{Unitarity}

A different type of limit on the SM parameters is obtained by
looking at high energy scattering.
For a
$2\rightarrow 2$ elastic scattering
process, 
the differential cross section is
\beq
{d \sigma \over d \Omega}={1\over 64 \pi^2 s} \mid {\cal A}\mid^2.
\eeq
Using a partial wave decomposition, the amplitude can be written as
\beq
{\cal A}=16 \pi \sum_{l=0}^{\infty} ( 2 l + 1)
P_l(\cos\theta)a_l   \, ,
\eeq
where $a_l$ is the spin $l$ partial wave and $P_l(\cos\theta)$ are
the Legendre polynomials.  The cross section
  is,
\beqn
\sigma &=& {8 \pi\over s}\sum_{l=0}^{\infty}
\sum_{l^\prime=0}^{\infty}
(2 l+1) (2 l^\prime + 1) a_l a^*_l
\int_{-1}^1 d \cos\theta P_l(\cos \theta) P_{l^\prime}(\cos \theta)
\nonumber \\
&=& {16 \pi \over s}\sum_{l=0}^{\infty}(2l+1)\mid a_l\mid^2 \, ,
\eeqn
where we have used the fact that the $P_l$'s are orthogonal.
The optical theorem  gives,
\beq
\sigma={1\over s}Im\biggl[{\cal A}(\theta=0)\biggr]
={16 \pi \over s}\sum_{l=0}^{\infty}(2l+1)\mid a_l\mid^2.
\eeq
This  immediately yields
 the unitarity requirement,
\beq
\mid a_l\mid^2=Im(a_l).
\eeq
or equivalently, 
\beq
\mid Re(a_l)\mid < {1\over 2}.
\eeq

As a demonstration of restrictions coming from the 
requirement of perturbative unitarity, consider the scattering
of longitudinal gauge bosons, $W^+_LW^-_L\rightarrow W^+_LW^-_L$,
The $J=0$ partial wave, $a_0^0$,  in the limit
$M_W^2<<s$, is\cite{Lee:1977eg,Duncan:1985vj},
\beqn
a_0^0(W_L^+W_L^-\rightarrow
W_L^+W_L^-)& \equiv &{1\over 16 \pi s}
\int^0_{-s}\mid {\cal A}\mid dt
\nonumber \\
&=&- {M_h^2\over 16 \pi v^2 }
\biggl[2 + {M_h^2\over s-M_h^2}-{M_h^2\over s}\log
\biggl(1+{s\over M_h^2}\biggr)\biggr].
\label{wwscat}
\eeqn
At very high energy, $s >>M_h^2$, Eq. ~\ref{wwscat}
has the limit
\beq
 a_0^0 \longrightarrow_{s>>M_h^2}
- {M_h^2\over 8 \pi v^2}.
\eeq
Applying the unitarity condition, $\mid Re (a_0^0)\mid< {1\over 2}$ gives
the restriction,
\beq
M_h< 870~GeV.
\label{hbound}
\eeq
It is important to understand that this does not mean that the
Higgs boson cannot be heavier than $870~GeV$, it simply means that
for heavier Higgs boson masses perturbation theory is  not valid.
The Higgs boson plays a fundamental role in the
theory since it cuts off the growth of the partial wave
amplitudes and makes the theory obey perturbative unitarity.

Taking the alternate limit, $s << M_h^2$, 
\beq
a_0^0 \longrightarrow_{s<<M_h^2}
-{s\over 32 \pi v^2}  \,  .
\eeq
Again applying the unitarity condition, we obtain,
\beq
\sqrt{s_c}< 1.7~TeV
\,  .
\label{units}
\eeq
The notation $s_c$ denotes $s$(critical), the scale
at which perturbative unitarity is violated.
Eq. ~\ref{units} is the basis for the oft-repeated statement,
{\it There must be new physics on the TeV scale}.  It is encouraging
that Eq. \ref{units} is exactly the energy scale that will be
explored at the LHC.

\section{Searches for the Higgs Boson at the LHC}

\begin{figure}
\includegraphics[scale=0.40,angle=-90]{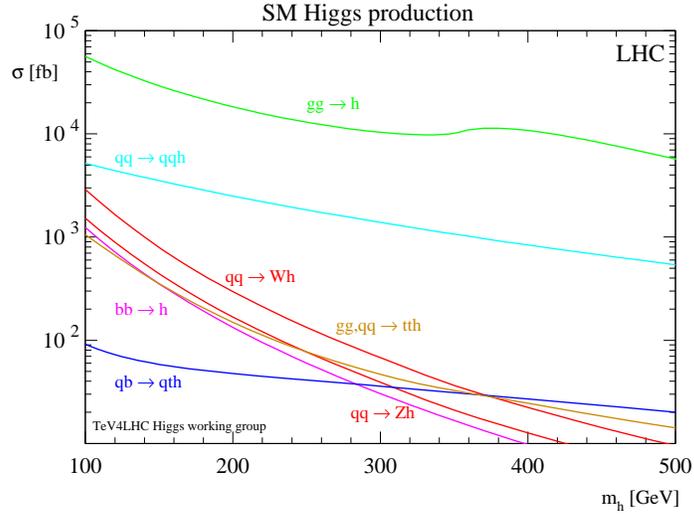}
\caption{Cross sections for SM Higgs boson production
processes at the LHC ($\sqrt{s}\!=\!14$~TeV), including
higher order corrections. From
Ref.~\cite{Aglietti:2006ne}.}
\label{fig:sm_higgs_lhc}
\end{figure}

\begin{figure}[t]
  \includegraphics[height=.15\textheight]{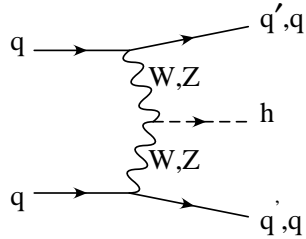}
  \caption{Higgs production from vector boson fusion.}
\label{fg:feynvbf}
\end{figure}

The LHC is expected to find the Higgs boson for all Higgs 
masses less than around $800~GeV$\cite{:1999fr,Ball:2007zza}.   The
production cross sections are large (Fig. \ref{fig:sm_higgs_lhc})
and the theoretical
predictions are well understood, with all important 
Higgs production channels known to at least
next-to-leading order accuracy\cite{:2008uu}.  As is the case at the Tevatron, 
the largest production mechanism is gluon fusion, but again
the largest decay for light Higgs bosons, 
$h\rightarrow b {\overline b}$, has an overwhelming  QCD
background. Above about $M_h\sim 140~GeV$,
the Higgs decays to $WW^*$ and $ZZ^*$ can be used for a Higgs discovery.

The vector
boson fusion channel (Fig. \ref{fg:feynvbf}), which is not important at the Tevatron,
is  useful for
Higgs discovery over a large Higgs 
mass region at the LHC\cite{Rainwater:2007cp}.
  By tagging the forward jets associated
with the Higgs production, 
the background can be significantly reduced. This channel can potentially
be used to observe
the decay $h\rightarrow \tau^+\tau^-$\cite{Plehn:1999xi} and $h\rightarrow
W^+W^-$\cite{Kauer:2000hi}.
 
\subsection{$h\rightarrow \gamma\gamma$}

Although the branching ratio is ${\cal O}(10^{-3}-10^{-4})$,
 (see Fig. \ref{fg:higgsbrs2}),
 the Higgs boson
can potentially be discovered in the $gg\rightarrow h\rightarrow \gamma\gamma$
channel for lighter Higgs bosons ($M_h < 140~GeV$).
   For $M_h >140~GeV$, the event rate becomes too small to be
observed.
  The largest reducible backgrounds are $q {\overline q}\rightarrow
\gamma\gamma$ and $gg\rightarrow \gamma\gamma$ which can be directly measured from
the sidebands away from the Higgs boson peak.  There are also large
reducible backgrounds 
from $\gamma $-jet and jet-jet production where the jet is misidentified
as a photon.  Excellent $\gamma$-jet separation and $\gamma$ energy resolution
help eliminate these backgrounds.  Both the ATLAS
and CMS collaborations have redone their original analyses to
optimize the event selection.  CMS finds that a
significance of $>8\sigma$ in the $h\rightarrow \gamma\gamma$ channel  can be 
achieved for $M_h\sim 130~GeV$ with an integrated luminosity
of $30~fb^{-1}$\cite{Goncalo:2008yg}, while the ATLAS studies are less
optimistic\cite{Goncalo:2008yg}.

\subsection{$h\rightarrow ZZ$}
\begin{figure}[t]
  \includegraphics[height=.3\textheight]{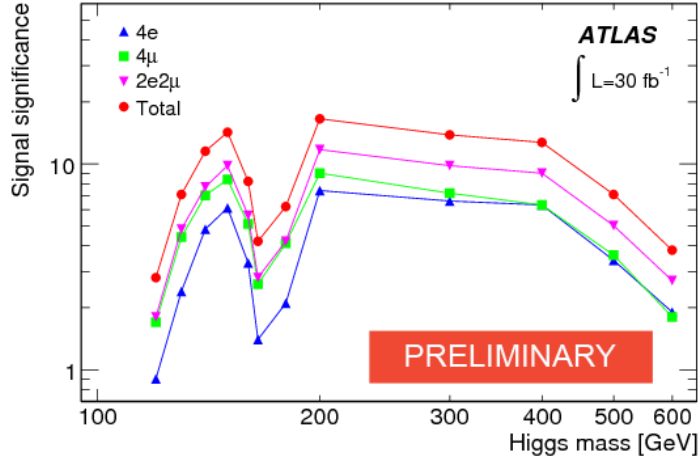}
  \caption{Significance of the $h\rightarrow ZZ^*
\rightarrow 4$~leptons discovery channel using the ATLAS
detector at the LHC with  $30~fb^{-1}$.  From Ref. \cite{Goncalo:2008yg}.}
\label{fg:4latlas}
\end{figure}

For $M_h > 130~GeV$, the Higgs boson can be discovered in the 
so-called golden channel,
$h\rightarrow ZZ^*\rightarrow 4$~leptons, except for near
the $W^+W^-$ threshold. This channel can be used for  Higgs masses
up to
around $M_h\sim 600~GeV$ and has  a smooth 
background   and a 
clean signature with a peak in the 4-lepton invariant mass allowing for
complete reconstruction of the Higgs mass.
 The background can be measured from the sidebands and
  the estimated sensitivity for $\int L= 30~fb^{-1}$ is
shown in Fig. \ref{fg:4latlas}.  

\subsection{Sensitivity}
\begin{figure}[t]
  \includegraphics[height=.35\textheight]{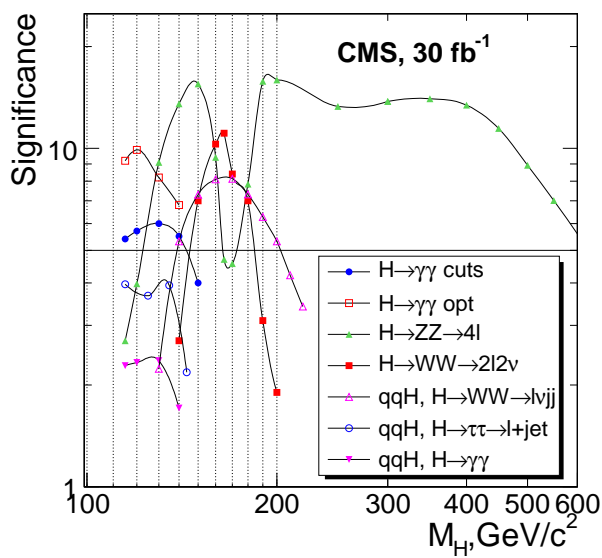}
  \caption{Significance of a Higgs discovery at CMS with $30~fb^{-1}$. 
From Ref. \cite{Ball:2007zza}.
    }
\label{fg:cmssens}
\end{figure}

The estimated sensitivity for a Higgs
discovery from the CMS experiment is shown in Fig. \ref{fg:cmssens}
for an integrated luminosity of  $\int L= 30~fb^{-1}$.  It is important to note 
that for any given Higgs mass, only a few channels are accessible and for
$M_h > 200 ~GeV$ only the $h\rightarrow ZZ\rightarrow 4$~leptons will be accessible
at the  initial luminosity.  The significance is greater than $5$ for all values
of the Higgs mass with $30~fb^{-1}$.
\begin{figure}[t]
  \includegraphics[height=.30\textheight]{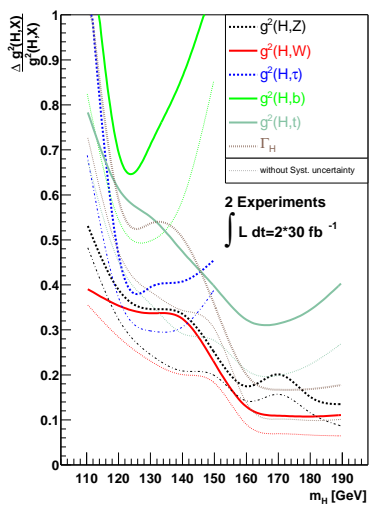}
  \caption{Potential sensitivity for Higgs coupling measurements
at the LHC. From Ref. \cite{Duhrssen:2004cv}.
    }
\label{fg:coups}
\end{figure}

After discovering the Higgs boson, the next task is to measure its properties as
precisely as possible to see if it is a SM Higgs boson.  We need to measure the
spin/parity, the Higgs couplings to fermions and gauge bosons, and the 
Higgs self-couplings. All
of these measurements will be extremely challenging at the LHC.  As an example, we show in
Fig. \ref{fg:coups} the precision which with the Higgs  couplings can
potentially be measured
at the LHC\cite{Duhrssen:2004cv,Belyaev:2002ua}. 

\section{Conclusion}
With the turn on of the LHC, particle physics will enter a new era of electroweak physics. 
There are three possibilities for the Higgs sector.  First, the Higgs could be discovered
with SM-like properties and a mass consistent with the electroweak precision observables.  In
this case, the problem of naturalness has no obvious solution and the only course will be
to measure the Higgs properties with great precision.  The second possibility is that
a Higgs boson is discovered with SM-like properties, but with a mass inconsistent with 
electroweak precision observables.  This case will keep theorists busy building models. 
Finally, it is possible that no Higgs boson will be discovered.  In this case, the problem
of unitarity comes to the forefront.  In all three cases, it is possible 
(and quite likely)
that new particles
outside the Higgs sector will be discovered.  Whatever the scenario, we are bound to
learn about the electroweak sector!

\begin{theacknowledgments}

Thanks to Maria Elena Tejeda-Yeomans, Alejandro Ayala, and the other organizers for a very enjoyable and interesting meeting. I thank the students for asking questions which made me think!
 This work was supported by the U.S. Department of Energy under grant
DE-AC02-98CH10886.

\end{theacknowledgments}

\bibliographystyle{aipproc}   

\bibliography{lecs}

\begin{thebibliography}{10}

\bibitem{Djouadi:2005gi}
A.~Djouadi,
\newblock Phys. Rept. {\bf 457}, 1 (2008), [hep-ph/0503172].

\bibitem{Quigg:2007fj}
C.~Quigg,
\newblock arXiv:0704.2045 [hep-ph].

\bibitem{Gunion:1989we}
J.~F. Gunion, H.~E. Haber, G.~L. Kane and S.~Dawson,
\newblock Cambridge, USA: Perseus Publishing (1990).

\bibitem{Peskin:1995ev}
M.~E. Peskin and D.~V. Schroeder,
\newblock Reading, USA: Addison-Wesley (1995) 842 p.

\bibitem{Rainwater:2007cp}
D.~Rainwater,
\newblock hep-ph/0702124.

\bibitem{Quigg:1983gw}
C.~Quigg,
\newblock Front. Phys. {\bf 56}, 1 (1983).

\bibitem{Dawson:1998yi}
S.~Dawson,
\newblock hep-ph/9901280.

\bibitem{Reina:2005ae}
L.~Reina,
\newblock hep-ph/0512377.

\bibitem{hdecay}
M.~Spira,
\newblock http://people.web.psi.ch/spira/proglist.html.

\bibitem{Frank:2006yh}
M.~Frank {\em et~al.},
\newblock JHEP {\bf 02}, 047 (2007), [hep-ph/0611326].

\bibitem{Aglietti:2006ne}
U.~Aglietti {\em et~al.},
\newblock hep-ph/0612172.

\bibitem{Barate:2003sz}
LEP Working Group for Higgs boson searches, R.~Barate {\em et~al.},
\newblock Phys. Lett. {\bf B565}, 61 (2003), [hep-ex/0306033].

\bibitem{Chang:2008cw}
S.~Chang, R.~Dermisek, J.~F. Gunion and N.~Weiner,
\newblock Ann. Rev. Nucl. Part. Sci. {\bf 58}, 75 (2008), [0801.4554].

\bibitem{Herndon:2008uv}
C.~Herndon, M. for the~Babar and D.~Collaborations,
\newblock 0810.3705.

\bibitem{Group:2008ds}
CDF, T.~T.~W. Group,
\newblock 0804.3423.

\bibitem{lepewwg}
{LEP Electroweak Working Group},
\newblock http://lepewwg.web.cern.ch/LEPEWWG/.

\bibitem{Amsler:2008zz}
Particle Data Group, C.~Amsler {\em et~al.},
\newblock Phys. Lett. {\bf B667}, 1 (2008).

\bibitem{Hollik:1988ii}
W.~F.~L. Hollik,
\newblock Fortschr. Phys. {\bf 38}, 165 (1990).

\bibitem{Jegerlehner:1991dq}
F.~Jegerlehner,
\newblock Lectures given at the Theoretical Advanced Study Institute in
  Elementary Particle Physics, (TASI), Boulder, Colo., Jun 3-29, 1990.

\bibitem{Sirlin:1980nh}
A.~Sirlin,
\newblock Phys. Rev. {\bf D22}, 971 (1980).

\bibitem{Appelquist:1974tg}
T.~Appelquist and J.~Carazzone,
\newblock Phys. Rev. {\bf D11}, 2856 (1975).

\bibitem{Flaecher:2008zq}
H.~Flaecher {\em et~al.},
\newblock 0811.0009.

\bibitem{Erler:2008ek}
J.~Erler and P.~Langacker,
\newblock 0807.3023.

\bibitem{Lindner:1985uk}
M.~Lindner,
\newblock Zeit. Phys. {\bf C31}, 295 (1986).

\bibitem{Chivukula:1999az}
R.~S. Chivukula and N.~J. Evans,
\newblock Phys. Lett. {\bf B464}, 244 (1999), [hep-ph/9907414].

\bibitem{Hambye:1996wb}
T.~Hambye and K.~Riesselmann,
\newblock Phys. Rev. {\bf D55}, 7255 (1997), [hep-ph/9610272].

\bibitem{Sher:1988mj}
M.~Sher,
\newblock Phys. Rept. {\bf 179}, 273 (1989).

\bibitem{Isidori:2001bm}
G.~Isidori, G.~Ridolfi and A.~Strumia,
\newblock Nucl. Phys. {\bf B609}, 387 (2001), [hep-ph/0104016].

\bibitem{Kolda:2000wi}
C.~F. Kolda and H.~Murayama,
\newblock JHEP {\bf 07}, 035 (2000), [hep-ph/0003170].

\bibitem{Drees:1996ca}
M.~Drees,
\newblock hep-ph/9611409.

\bibitem{Dawson:1997tz}
S.~Dawson,
\newblock hep-ph/9712464.

\bibitem{Lee:1977eg}
B.~W. Lee, C.~Quigg and H.~B. Thacker,
\newblock Phys. Rev. {\bf D16}, 1519 (1977).

\bibitem{Duncan:1985vj}
M.~J. Duncan, G.~L. Kane and W.~W. Repko,
\newblock Nucl. Phys. {\bf B272}, 517 (1986).

\bibitem{:1999fr}
CERN-LHCC-99-15.

\bibitem{Ball:2007zza}
CMS, G.~L. Bayatian {\em et~al.},
\newblock J. Phys. {\bf G34}, 995 (2007).

\bibitem{:2008uu}
N.~E. Adam {\em et~al.},
\newblock 0803.1154.

\bibitem{Plehn:1999xi}
T.~Plehn, D.~L. Rainwater and D.~Zeppenfeld,
\newblock Phys. Rev. {\bf D61}, 093005 (2000), [hep-ph/9911385].

\bibitem{Kauer:2000hi}
N.~Kauer, T.~Plehn, D.~L. Rainwater and D.~Zeppenfeld,
\newblock Phys. Lett. {\bf B503}, 113 (2001), [hep-ph/0012351].

\bibitem{Goncalo:2008yg}
R.~Goncalo, f.~t. ATLAS and C.~Collaborations,
\newblock 0811.3778.

\bibitem{Duhrssen:2004cv}
M.~Duhrssen {\em et~al.},
\newblock Phys. Rev. {\bf D70}, 113009 (2004), [hep-ph/0406323].

\bibitem{Belyaev:2002ua}
A.~Belyaev and L.~Reina,
\newblock JHEP {\bf 08}, 041 (2002), [hep-ph/0205270].

\end{thebibliography}

\end{document}